\documentclass[submission,copyright,creativecommons]{eptcs}
\providecommand{\event}{Post Proceedings of ADG'23} 
\usepackage[T1]{fontenc}
\usepackage{graphicx}
\usepackage{doi}

\usepackage{xurl,amsmath}

\begin{document}
\title{Showing Proofs, Assessing Difficulty \\ with GeoGebra Discovery\thanks{Authors supported by a grant PID2020-113192GB-I00 (Mathematical Visualization: Foundations, Algorithms and Applications) from the Spanish MICINN.}}
\def\titlerunning{Showing Proofs, Assessing Difficulty with GeoGebra Discovery}
%
%
%
\def\authorrunning{Z.~Kov\'acs et al.}
%
\author{
Zolt\'an Kov\'acs
\institute{The Private University College of Education of the Diocese of Linz, Austria}
\email{zoltan.kovacs@ph-linz.at}
\and
Tom\'as Recio
\institute{Escuela Polit\'ecnica Superior, Universidad Antonio de Nebrija, Madrid, Spain}
\email{trecio@nebrija.es}
\and
M. Pilar V\'elez
\institute{Escuela Polit\'ecnica Superior, Universidad Antonio de Nebrija, Madrid, Spain}
\email{pvelez@nebrija.es}}
\maketitle              
\begin{abstract}
In our contribution we describe some on-going improvements concerning the Automated Reasoning Tools developed in GeoGebra Discovery, providing different examples of the performance of these new features.  
We describe the new {\tt ShowProof} command, that outputs both the sequence of the different steps performed by GeoGebra Discovery to  confirm a certain statement, as well as a number intending to  grade the difficulty or interest of the assertion. The proposal of this assessment measure, involving the comparison of the expression of the thesis (or  conclusion) as a combination of the hypotheses, will be developed.

\end{abstract}

\section{Introduction}
In the past years we have been developing and including, both in the standard version of GeoGebra (GG) as well as in the fork version GeoGebra
Discovery,\footnote{See project page \url{https://github.com/kovzol/geogebra-discovery}.} different automated reasoning tools. See \cite{KRV22} for a general description and references.

The goal of the current contribution is to present some ongoing work regarding two different, but related, important improvements of GeoGebra Discovery. One, to visualize the different steps that GG Discovery performs with a given geometric statement until it declares its truth (or failure). Two, to test, through different elementary examples, the suitability of an original proposal to evaluate the interest, complexity or difficulty of a given statement. Let us advance that our proposal involves the notion of syzygy of a set of polynomials.

The relevance of showing details about each of the steps performed by our automated reasoning algorithms implemented in GG Discovery is quite evident. In fact, as a consequence of the result in~\cite{BBN}, describing the formalization of the arithmetization of Euclidean plane geometry, proofs of geometric statements obtained using algebraic geometry algorithms are also valid on the synthetic geometry realm. Although it might seem obvious that synthetic proofs  facilitate human understanding of a geometric statement, as compared with the difficult interpretation associated to algebraic proofs, this assertion could be a matter of discussion if considered for statements (and for human minds) of a certain level, a discussion that it is out of the scope of the present paper, and that could be the subject of a future research,  in an educational context. 

On the other hand, the evaluation of the {\it difficulty} of geometric statements is an old subject, regardless of its relation to automated proving. We can mention the work of Lemoine on the number of steps required by a geometric construction (a higher number of steps =  a more complicated construction), a proposal that has been thoroughly studied, adapted to the Dynamic Geometry context, and exemplified in different repositories, e.g., in several recent works by Quaresma and collaborators, such  as \cite{QSGB,SBQ}. 

But the complexity of a geometric construction is not, in general,  a good measure to establish the difficulty of a statement involving the same construction: one can make a very complicated figure and then conjecture some obvious property, easy to derive from the construction steps.

Thus, in this paper we make a preliminary report of our current work aiming to establish some {\it difficulty} criteria,  that profits from our algebraic approach to proving geometric statements. Roughly: we propose to label as {\it more difficult} those statements where:
\begin{enumerate}
\item first of all,   the polynomial involved in the description of the thesis  (or conclusion)  is a sum of products of  the hypotheses polynomials multiplied by  some other polynomials  (thus, the statement is just formally {\it true}). These other polynomials are usually named as  syzygies,
\item else, there is a combination of $1$  as a sum of products of  the hypotheses polynomials and of the negation of the thesis $T$ (expressed as $T\cdot t-1$) multiplied by  some other polynomials  (thus, the statement is, by {\it reductio ad absurdum}, {\it geometrically true}, meaning that a power of the thesis is a combination of the hypotheses).  Again, these other polynomials that multiply the negation of the thesis and the hypotheses, are usually named as syzygies,
\item these expressions of the thesis (or of $1$) as a combination of the hypotheses (of the hypotheses and the negation of the thesis) requires higher degree polynomials, i.e.~more complicated syzygies.
\end{enumerate}

 Of course, although we are aware that {\it difficulty} and {\it interest} are not identical concepts, let us recall that  this notion, the {\it interestingness} of theorems, is also subject to current research (see \cite{GGC,GLC,PGS,CBW}) in different contexts (A.I., Big Data, etc.), sometimes explicitly referring to our particular automated reasoning  in geometry  approach. 

In what follows we will describe, mostly through some examples, our ongoing work on these two subjects.

\section{ShowProof Command}

Unfortunately, till now, the algebraic geometry nature of the algorithms behind the automatic reasoning tools  implemented in GeoGebra or GeoGebra Discovery do not allow providing readable arguments justifying their outputs. Computations are performed in the background, using some embedded Computer Algebra System, such as Giac \cite{GiacGG-RICAM2013}. The user only gets a kind of yes/no answer.

The {\tt ShowProof} command is a first attempt to enhance the visibility of the proofs achieved by GeoGebra Discovery, by showing the result of the different steps performed by GeoGebra Discovery to  confirm a certain statement: algebraic translation of the geometric input construction, numerical specialization of the coordinates of  some free points, automatic inclusion of non-degeneracy conditions,  and writing -- using the concept of syzygy and its computation --   the expression of $1$ as a combination of the hypotheses and  the negation of the thesis (thus proving the statement by {\it reductio ad absurdum}), or of the thesis as a combination of the hypotheses (direct proof of the statement).

Visualizing the output of most of these steps is just a question of improving the user interface, as it does not involve any new computation. This is so except for the (most important) last two items: the concrete expression of 1 (or of the thesis) as a combination of other polynomials. Indeed, to decide that a statement is generally true \cite{KRV22}  GeoGebra Discovery just has to perform some elimination of the ideal of hypotheses  plus the negation of the thesis,  and to verify that it is not zero. Then, adding to the hypotheses ideal  the negation of some of the generators $g$ of this elimination (using Rabinowitsch trick, as $g\cdot t-1$), it is obvious that the (new hypotheses + negation of thesis) ideal contains $1$, since it will include both $g$ and $g\cdot t-1$.

But the user, who has not  any concrete evidence about the result of the previous elimination (for example, expressing the added non-degeneracy condition as a combination of the (given hypotheses + negation of thesis) ideal, probably would appreciate
handling an expression of 1 as a combination of the (new hypotheses + negation of thesis) ideal.  Or, even more impacting, viewing the thesis as a combination of the hypotheses. Notice that the first possibility just means that {\it over the complexes} the thesis vanishes over all points satisfying the new hypotheses (the given ones and the added non-degeneracy condition) and, thus, that a power of the thesis is a combination of the hypotheses. We can say that the statement is, in this case, {\it geometrically true}, while, if the thesis itself is a combination of the hypotheses, we can declare we have a {\it formally true} statement.

Both issues are now addressed through the  {\tt ShowProof} command, using the concept of syzygy (e.g.~\cite {GP}, page 104):

 Given any $G = (g_1,\ldots,g_s) \in k[x_1,\ldots,x_n]^s$, we can define a syzygy on $G$ to be an $s$-tuple $S = (h_1,\ldots,h_s) \in k[x_1,\ldots, x_n]^s$,  such that $\sum _i  h_i\cdot g_i=0$.

In particular, if we include $1$ (respectively, the thesis) as the first element of the collection of polynomials $G$, and the remaining elements are the generators of the ideal of the new hypotheses  plus the negation of the thesis (respectively, of the new hypotheses), then we will get (if the statement is geometrically true, respectively, symbolically true) syzygies of the kind $(1,-h_2, \dots,-h_s)$, allowing us to output a concrete expression of $1$ (respectively, the thesis) as a combination of the  new hypotheses  plus the negation of the thesis (respectively, of the new hypotheses).

\begin{figure}
\begin{center}\includegraphics[scale=0.60]{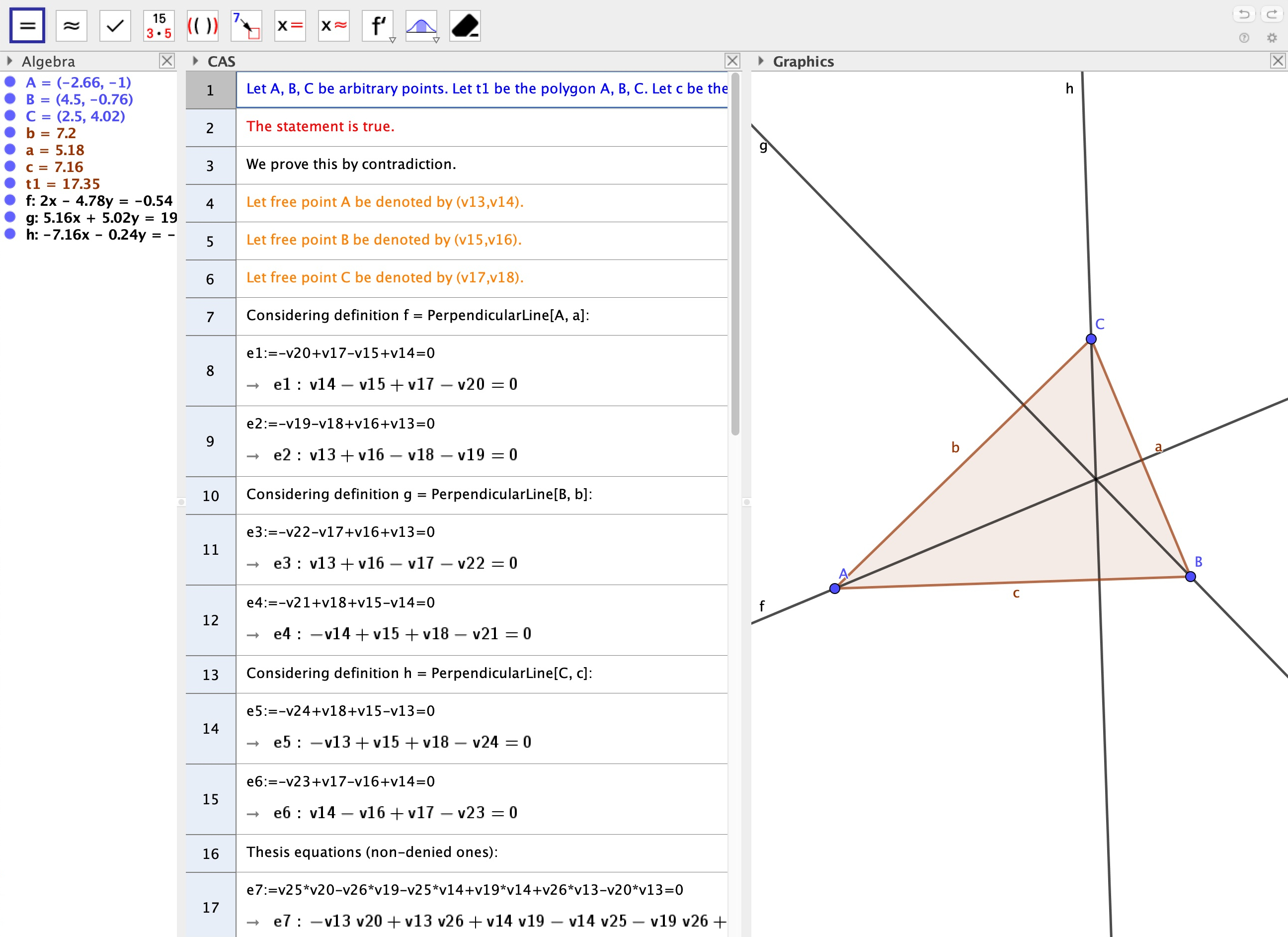}
\caption{Viewing proof of the {\it Intersection of heights} theorem through {\tt ShowProof}. Initial steps. } \label{fig1}
\end{center}
\end{figure}

The next figures illustrate the current output of  the {\tt ShowProof} command. Figure \ref{fig1} displays, first,  the automatically and internally assigned coordinates of the free vertices of the triangle $A, B, C$; then, the equations of the different heights; finally, the thesis (the fact that the last height goes through the intersection of the other two).

\begin{figure}
\begin{center}
\includegraphics[scale=0.60]{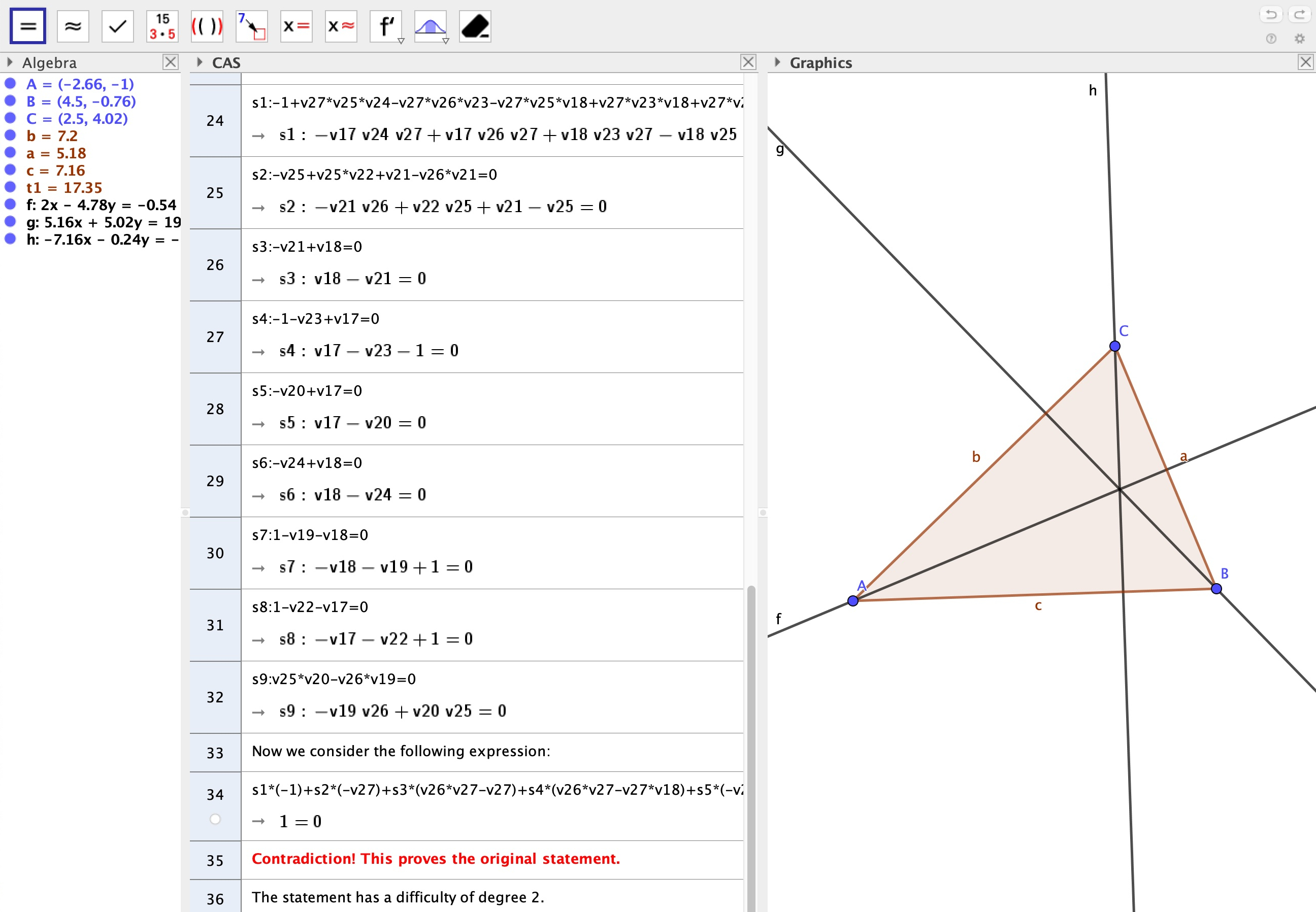}
\caption{Viewing proof of the {\it Intersection of heights} theorem through {\tt ShowProof}. Specialized equations, conclusion and difficulty. } \label{fig2}
\end{center}
\end{figure}

Next, following our algorithm and without loss of generality, the program automatically specializes the given free coordinates, to reduce the number of variables before starting the computations. This is shown
in figure \ref{fig2}, that ends declaring that the statement is {\it geometrically true} by explicitly showing $1$  as a combination of the negation of the thesis and the hypotheses equations (thus, $1=0$, since all these equations are equal to zero). Finally, the last line declares that this statement is of difficulty 2, a measure that we will roughly describe in the next section.

\section{Interestingness}
Although the precise formulation of the following reflections require a serious and future research analysis, that is not the goal of the present paper,  we dare to consider quite evident that  showing arguments for the truth of a geometric statement is important in the scientific and educational context, even more relevant nowadays, in a context of close collaboration (or leadership?) of machine-driven learning. Analogously, under the dominance and ubiquitousness of machine learning environments,  very prone to automatically produce large amounts of geometric information,  we think it is very relevant to develop instruments that allow humans to assess the relevance of the obtained results.

\begin{figure}
\begin{center}
\includegraphics[width=0.9\textwidth]{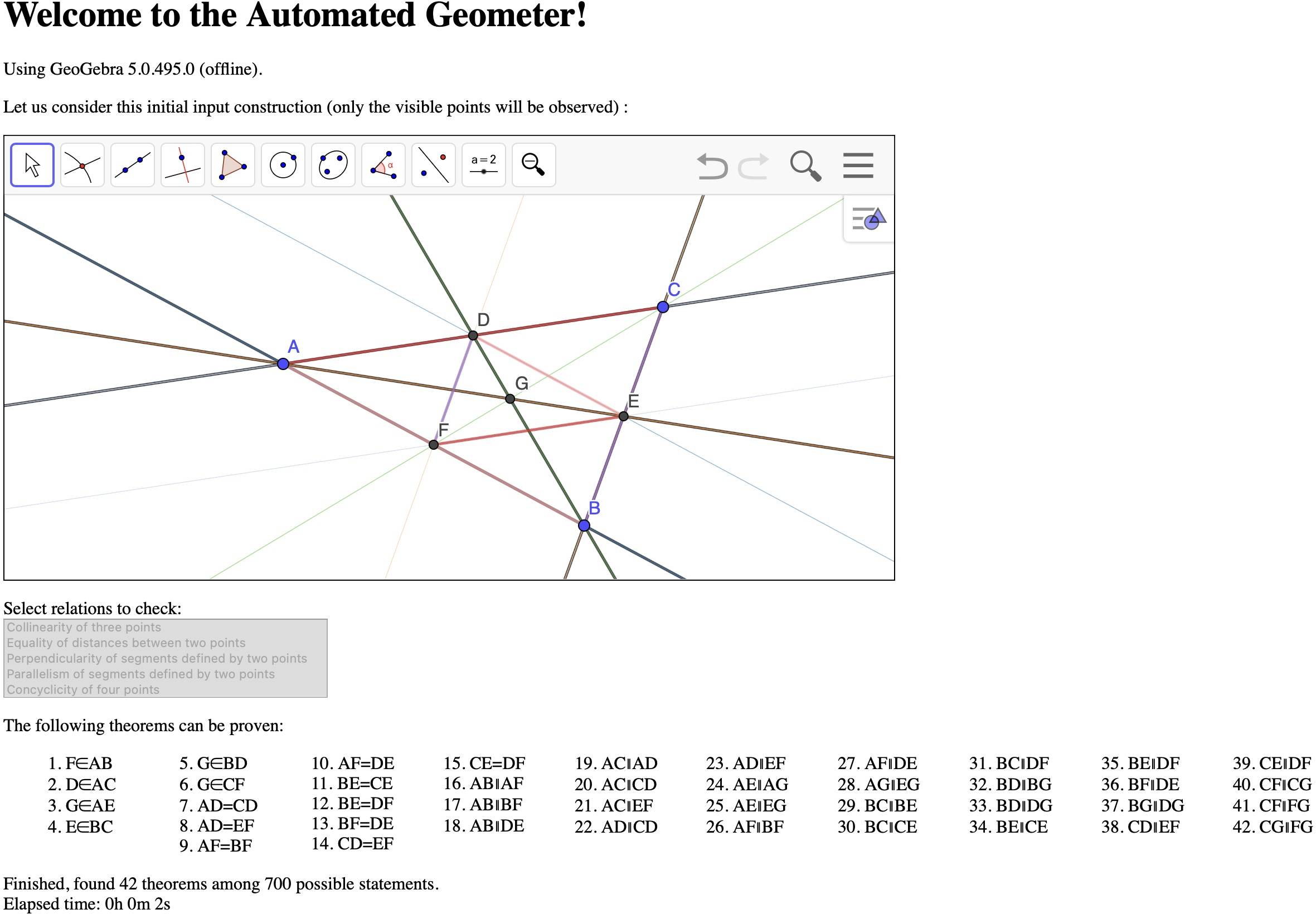}
\caption{Automated Geometer: relevant (e.g.~Theorem 7,  medians intersect at a common point) and trivial (e.g.~Theorem 1, midpoint $F$  of $AB$ in the line $AB$, or Theorem 9, midpoint $F$ is equidistant from $A$ and $B$) results. } \label{fig3}
\end{center}
\end{figure}

For example, GeoGebra Discovery has already a {\it Discover} command that automatically finds all statements of a certain kind that hold over some element of a construction (and a more general command in a web version, the
{\it Automated Geometer},\footnote{\url {http://autgeo.online/ag/automated-geometer.html?offline=1}}
automatically finding {\it all} the statements of a selected kind (collinearity of three points, equality of distances between two points, etc., as declared in the dark box below the geometric figure in figure \ref{fig3})  holding over {\it all} elements of a figure.) See also the recent {\tt StepwiseDiscovery} command, that discovers automatically all statements involving each of the new elements that the user is adding in each construction step.  Now, it happens that a great number of such {\it discovered} statements are just obvious! See figure \ref{fig3}, for some examples.

Thus, we dare to introduce, as a measure of the complexity of a result, the following definition:

We say that a statement $H \Rightarrow T$ is of complexity $d$ if $d$ is the maximum degree of the syzigies expressing $T$ (or 1) as a combination of the hypotheses (correspondingly, of the hypotheses and the negation of the thesis).

In what follows we will describe different examples, towards analyzing the potential of  this proposal for the concept of  an {\it interesting} statement. 

\subsection{Example 1}

Let us build a point  $F$ as the midpoint of $AB$ (see Figure \ref{fig3}), and ask about {\it Discover(F)}.  The program {\it discovers} (among other, really interesting statements),  that  the length of segment $FA$ is equal to the length of $FB$. But, since the definition of $F$ as midpoint of the segment with extremes  $A(a_1,a_2), B(b_1,b_2)$  is that the coordinates of $F$ are $((a_1+b_1)/2, (a_2+b_2)/2)$, it immediately follows that  $FA=((a_1-b_1)/2, (a_2-b_2)/2)$ and $FB=(b_1-a_1)/2, (b_2-a_2)/2)$. Obviously, from these coordinates, it follows that the length of both segments is identical, it only requires to simplify $((a_1-b_1)/2)^2+((a_2-b_2)/2)^2-(((b_1-a_1)/2)^2+((b_2-a_2)/2)^2)$,
yielding 0. Thus, we could roughly declare that this thesis is just the trivial equation $0=0$ and thus it is always a combination of whatever set of hypotheses multiplied by zero. Thus the degree of the zero polynomial $0$ (that some algebraist consider as a negative number) could be a measure of the complexity of this highly trivial statement.

\subsection{Example 2}
On a different example about elements in the same figure, if we consider the coordinates of point $G(g_1,g_2)$, the intersection of the line that goes from $A$ to the midpoint $E(e_1, e_2)$ of side $BC$, and of the line that goes from $B$ to the midpoint $D(d_1,d_2)$ of side $AC$, where $C(c_1,c_2)$,  we have:

\begin{itemize}
\item
Coordinates of $E, D$:
 \begin{equation}
H_1, H_2:   e_1=(b_1+c_1)/2, e_2=(b_2+c_2)/2,
\end{equation}
\begin{equation}
H_3, H_4: d_1=(a_1+c_1)/2, d_2=(a_2+c_2)/2.
\end{equation}

\item
Equations of lines $A,E$ and $B,D$
\begin{equation}
\textrm{line } AE: (x-a_1)\cdot(e_2-a_2) -(y-a_2)\cdot(e_1-a_1)=0,
\end{equation}
\begin{equation}
 \textrm{line } BD: (x-b_1)\cdot(d_2-b_2) -(y-b_2)\cdot(d_1-b_1)=0.
\end{equation}

\item Thus, when we declare $G$ as the intersection of these two lines, we introduce the equations

\begin{equation}
H_5: \textrm{ line } AE: (g_1-a_1)\cdot(e_2-a_2) -(g_2-a_2)\cdot(e_1-a_1)=0,
\end{equation}
\begin{equation}
H_6: \textrm{ line } BD: (g_1-b_1)\cdot(d_2-b_2) -(g_2-b_2)\cdot(d_1-b_1)=0.
\end{equation}

\item Moreover, let us recall that midpoint $F$ of  side $AB$ has coordinates:
\begin{equation}
H_7, H_8: f_1=(a_1+b_1)/2, f_2=(a_2+b_2)/2.
\end{equation}

\item Finally, the Theorem 6 in figure \ref{fig3}, namely, $G \in CF$,  means that
\begin{equation}
\textit{Thesis: } G \in \textrm{ line } CF: (g_1-c_1)\cdot(f_2-c_2) -(g_2-c_2)\cdot(f_1-c_1)=0.
\end{equation}
\end{itemize}

Now, after some involved computations on the syzygies of the set
$$(\textit{Thesis}, H_1, H_2, \dots, H_8),$$ here we use the same
notation for the \textit{Thesis} and the hypotheses $H_i$, but
considering just the involved polynomials, not the equality to zero. We
find that we can express the \textit{Thesis} as a combination of $(H_1,
H_2, \dots,H_8)$ multiplied by linear polynomials in the involved
variables $$a_1,a_2,b_1,b_2,c_1,c_2,d_1,d_2,e_1,e_2,f_1,f_2,g_1,g_2,$$
namely, the thesis is equal to
\begin{equation}
\textit{Thesis}=(-g_1 + f_1)\cdot G[1]+(-f_2 + g_2)\cdot G[2]+(1)\cdot G[7]
\end{equation}
where  $G[1],G[2],G[7]$ are elements of the Gr\"obner  basis of the hypotheses ideal with respect to the \textit{plex} order, and can be expressed, respectively, as sums of products of  $[0, 0, 0, -1/2, 0, -1/2, 0, 1/2]$,  $[0, 0, -1/2, 0, -1/2, 0, 1/2, 0]$, $[0, 2\cdot g_2 - 2, -2\cdot g_1, -2\cdot g_2, 0, 2, 2\cdot g_1, -2]$,  times $[H_1, H_2, \dots, H_8]$.

In summary, the thesis is a combination of the hypotheses multiplied by polynomials of degree at most 1. This degree 1, vs.~the negative degree of the syzygies in the previous statement shows, it is our proposal, that this second statement, about the coincidence of the intersection of the three medians of a triangle, is {\it more difficult} than the statement about the equality of lengths of the two segments determined by the mid-point.

Let us remark that this same statement,  but specializing vertices $A=(0,0)$, $B=(0,1)$, leads to degree 2 syzygies
\begin{align}
\textit{Thesis}=(-c_1/2 + g_1/2)\cdot G[1]&+( -g_2 + c_2)\cdot G[2]+( -g_1)\cdot G[6] \nonumber\\
&+(g_2 - 1/2)\cdot G[7] -1\cdot G[8]
\end{align}
where, again, $G[1],G[2],G[6],G[7],G[8]$ are elements of the Gr\"obner
basis of the (specialized) hypotheses ideal with respect to the
\textit{tdeg} order, and can be expressed, respectively, as sums of
products of  $[0, 0, 0, 0, 2, 0, 0, 0]$,  $[1, 0, 0, 0, 0, 0, 0, 0]$, $[0,
0, 0, 0, 0, 0, -2, 0]$, $[0, 2\cdot g_2 - 2, -2\cdot g_1, -2\cdot g_2, 0,
2, 2\cdot g_1, -2]$, $[0, 2\cdot (g_2 - 1)\cdot g_2, -2\cdot g_1\cdot g_2,
-2\cdot (g_2 - 1)\cdot g_2, 0, 2\cdot g_2, 2\cdot g_1\cdot g_2, -2\cdot
g_2 + 2]$,  times $[H_1, H_2, \dots, H_8]$, notice that only the summand
involving $G[7]$ rises the degree to degree two.

\subsection{Example 3}

\tolerance10000

As a third example, let us start (see Figure \ref{fig1}) with a triangle with vertices $A(v13, v14)$, $B(v15, v16)$, $C(v17,v18)$. Then, consider the perpendicular line through $A$ to side $BC$.  Considering $P(v19, v20)$ as the coordinates of a generic point $P$ in this line, we obtain the equation 
\begin{equation}
H_1: (v19-v13) \cdot(v17-v15)   + (v20-v14)\cdot(v18-v16) =0.
\end{equation}
Similarly for the other two heights.

Thus, the statement {\it the three heights of a triangle have a common intersection} is a matter of considering two hypotheses (the equation of height from $A$ and from, say, $B$) with a common generic point $P$. And the thesis is also one equation, namely, showing that this generic point $P$ satisfies that line $PC$ is perpendicular to line $AB$. More precisely:

\begin{itemize}
 \item Hypotheses:
\begin{itemize}
\item Height from $A$:
\begin{equation}
 (v19-v13) \cdot(v17-v15)   + (v20-v14)\cdot(v18-v16)  =0.
\end{equation}
\item Height from $B$:
\begin{equation}
(v19-v15) \cdot(v17-v13)   + (v20-v16)\cdot(v18-v14)  =0.
\end{equation}
\end{itemize}
\item Thesis:
\begin{equation}
(v19-v17)\cdot(v15-v13)   + (v20-v18)\cdot(v16-v14)  =0.
\end{equation}
\end{itemize}

 Notice that these equations seem different from those describing this statement in figure \ref{fig1}. There, the height from $A$ is described by considering the coordinates of  point $X(v19,v20)$ in this line, and regarding such point as the translation of $A$ by a vector $ AX$ perpendicular to $BC$ so that $-v20+v17-v15+v14=0, -v19-v18+v16+v13=0$. Finally, Equation (17) in figure \ref{fig1}, describes a generic point $(v25,v26)$ of this line, verifying  $v25\cdot v20-v26\cdot v19-v25\cdot v14+v19\cdot v14+v26\cdot v13-v20\cdot v13=0$.

 But let us remark that this equation is practically the same as the one above introduced for the height from $A$, replacing there generic point coordinates $(v19, v20)$ by  $(v25,v26)$ and,  in the new equation,  $v20=v17-v15+v14, v19=v16+v13-v18$,  
yielding
  \begin{align*}
 &\;v25\cdot v20-v26\cdot v19-v25\cdot v14+v19\cdot v14+v26\cdot v13-v20\cdot v13=\\
 =&\;v25\cdot (v17-v15+v14)-v26\cdot (v16+v13-v18)-v25\cdot v14+\\
 &\;(v16+v13-v18)\cdot v14+v26\cdot v13-(v17-v15+v14)\cdot v13=\\
 = &\;(v15 - v17)\cdot (v13-v25)+(v16-v18)\cdot (v14-v26).
  \end{align*}

Similarly, the thesis is now, with the notation of figure \ref{fig1}:
\begin{equation}
v25\cdot v24-v26\cdot v23-v25\cdot v18+v23\cdot v18+v26\cdot v17-v24\cdot v17 =0
\end{equation}
that belongs to the new ideal of  hypotheses
$\langle-v20+v17-v15+v14,-v19-v18+v16+v13,-v22-v17+v16+v13,
-v21+v18+v15-v14, -v24+v18+v15-v13, -v23+v17-v16+v14,
v25\cdot v20-v26\cdot v19-v25\cdot v14+v19\cdot v14+v26\cdot v13-v20\cdot v13,
v25\cdot v22-v26\cdot v21-v25\cdot v16+v21\cdot v16+v26\cdot v15-v22\cdot v15\rangle.$

The interesting point here is that the {\it complexity} of the statement is different if one considers the previous equations or the ones in figure \ref{fig1}. In the first case, it is immediate to see that the thesis is the difference of the height from $B$ minus the height from $A$. The same result holds if we do specialize numerically $A, B$,  as it is usual in GeoGebra Discovery, considering $A(0,0), B(0,1)$, so that the above hypotheses turn out to be:
\begin{itemize}
\item Height from $A$:
\begin{equation}
 (v19-0) \cdot (v17-0)   + (v20-0)\cdot (v18-1)  =0.
\end{equation}
\item Height from $B$:
\begin{equation}
(v19-0) \cdot (v17-0)   + (v20-1)\cdot (v18-0)  =0.
\end{equation}
\item Thesis:
\begin{equation}
(v19-v17) \cdot (0-0)   + (v20-v18)\cdot (1-0)  =0.
\end{equation}
\end{itemize}

 So we could say that here the syzygies are of degree 0, and thus we can declare the complexity of the statement is 0, as  the thesis is just a linear combination of the hypotheses (multiplied by constants).

But, with the equations in figure \ref{fig1}, using the same \textit{tdeg} order, one gets the thesis as a  combination of the hypothesis  multiplied by polynomials of degree at most one, so the complexity rises one element and, if we do specialize $A(0,0), B(0,1)$ in the equations of figure \ref{fig1}, we get that the syzygies have to be of degree 2, as stated in figure \ref{fig2}!

 \section{Conclusions}
The explicit visualization of the algebraic expression that connects hypotheses and thesis (or expresses number 1 as a combination of hypotheses and the negation of the thesis) seems a relevant improvement of GeoGebra Discovery automated reasoning tools, allowing the user not only to be able to personally confirm the truth of the given statement, but also to measure its difficulty. 

Of course, the proposal of this measure is just on its initial steps and requires much further research. For example, extending the already considered examples, including others of higher difficulty, such as those concerning Mathematics Olympiad problems. Indeed, we have already verified  with some of them that our measure of the computed difficulty agrees that they are much more difficult than the traditional school problems, but analyzing more problems from different sources, comparing our rank and the behavior of the olympic teams on the same problems, is still on-going work \cite{ARP}.

Moreover, the proposed research should  not just be restricted to producing benchmarks, but to reflect on several more conceptual issues, such as the role, on the measure of the complexity, of the different ways to express the same statement (as remarked in the previous examples). Ditto, for the specialization of some variables, for choosing different ordering for computing syzygies, etc.  We also need to understand  the difference, or the connection, between the complexity of expressing a statement and the complexity of deciding its truth, the relation between the complexity of the statement and the complexity of the associated ideal membership problem, etc.

Indeed, we think there is plenty of work in this context and it is our intention to address such issues in a near future.


%
%

%
%
%
%

\end{document}